\documentclass[twocolumn,secnumarabic,amssymb, nobibnotes, aps, prd,a4paper,superscriptaddress,showpacs]{revtex4-1}

\usepackage{siunitx}
\usepackage{amsmath,amssymb,array,booktabs,graphicx,lmodern,latexsym}
\usepackage{color,wasysym,varioref,xspace,txfonts,dcolumn,framed}

\usepackage{hyperref}
\usepackage{breakurl}

\newcommand{\fref}[1]{FIG.~\ref{#1}}

\newcommand{\tref}[1]{TABLE.~\ref{#1}}

\newcommand{\radl}{X_0}

\newlength{\collength}
\newlength{\dcollength}

\newcommand{\CERN}{\text{CERN}}

\newcommand{\lmult}{\ell_\gamma}

\usepackage{color,hyperref}
\definecolor{darkblue}{rgb}{0.0,0.0,0.3}
\hypersetup{colorlinks,breaklinks,
            linkcolor=darkblue,urlcolor=darkblue,
            anchorcolor=darkblue,citecolor=darkblue}

\begin{document}
%

\title{Experimental investigation of the Landau-Pomeranchuk-Migdal effect in low-$Z$ targets}
\author{K.K. Andersen}
\affiliation{Department of Physics and Astronomy, Aarhus University, Denmark}
\author{S.L. Andersen}
\affiliation{Department of Physics and Astronomy, Aarhus University, Denmark}
\author{J. Esberg}
\affiliation{CERN, Geneva, Switzerland}
\author{H. Knudsen}
\affiliation{Department of Physics and Astronomy, Aarhus University, Denmark}
\author{R.E. Mikkelsen}
\affiliation{Department of Physics and Astronomy, Aarhus University, Denmark}
\author{U.I. Uggerh{\o}j}
\email{ulrik@phys.au.dk}
\affiliation{Department of Physics and Astronomy, Aarhus University, Denmark}
\author{T.N. Wistisen}
\affiliation{Department of Physics and Astronomy, Aarhus University, Denmark}
\author{P. Sona}
\affiliation{University of Florence, Florence, Italy}
\author{A. Mangiarotti}
\affiliation{University of S\~{a}o Paulo, Brazil}
\author{T.J. Ketel}
\affiliation{Free University, Amsterdam, The Netherlands}



\collaboration{CERN NA63}

\date{\today}

\begin{abstract}
In the \CERN{} NA63 collaboration we have addressed the question of the potential inadequacy of the commonly used Migdal formulation of the Landau-Pomeranchuk-Migdal (LPM) effect by measuring the photon emission by \num{20} and \SI{178}{GeV} electrons in the range 100 MeV - 4 GeV, in targets of LowDensityPolyEthylene (LDPE), C, Al, Ti, Fe, Cu, Mo and, as a reference target, Ta. For each target and energy, a comparison between si\-mu\-lated values based on the LPM suppression of incoherent bremsstrahlung is shown, taking multi-photon effects into account. For these targets and energies, we find that Migdal's theoretical formulation is adequate to a precision of better than about 5\%, irrespective of the target substance.
\end{abstract}


\maketitle


\section{Introduction}
The well-known Landau-Pomeranchuk-Migdal  (LPM) effect, which reduces the radiation emission from an ultrarelativistic electron in a medium due to multiple Coulomb scattering, has been investigated extensively at both SLAC and CERN. Nevertheless, the question remains if the 'traditional' approach of using Migdal's theoretical formulation is adequate in the case of targets of low nuclear charge. Since the LPM effect is routinely applied in codes used to interpret extended air showers which naturally occur in a low-$Z$ medium, the answer to this question is important. There is reason to suspect a certain degree of inadequacy since Migdal applied the Thomas-Fermi approach to determine the screening \cite[eq.\ (22)]{Migd56}, an approach that is inherently statistical and therefore inaccurate for atoms with few electrons, see e.g.\ \cite{Gomb56}. As essentially all experiments are performed in the full-screening limit, the inaccuracy of the screening model might lead to imprecise answers. Furthermore, the contribution from electrons may be influenced differently by the LPM effect than the nuclear contribution since a different range of momentum transfers are relevant and different screening parameters involved \cite{Whee39}. And, finally, seemingly unphysical 'kinks' in the calculated spectra based on Migdal's formulation appear when the Migdal formula reaches the Bethe-Heitler (BH) level.

Numerous previous experiments have presented evidence for the LPM effect, see e.g.\ \cite{Klei99}, in particular the SLAC experiment performed with 8 and 25 GeV electrons \cite{Anth95,Anth97}, and the experiments at energies up to 287 GeV performed at CERN \cite{Hans03,Hans04}.

\section{Experiment}

The present experiment was performed in the H4 beam line in the North Experimental Area of the \CERN{} Super Proton Synchrotron with
tertiary beams of electrons with energies of 20 and 178 GeV. The 20 GeV beam was chosen to investigate experimentally the 'kink' in the radiation spectrum obtained from Migdal's formula. The high energy was chosen to maximize the photon formation length, given by $l_f = 2\gamma^2c/\omega$, associated to the emission of photons with energies in the sensitive range of the Bismuth Germanate (BGO) calorimeter used, while retaining an acceptable beam intensity. 

\begin{figure*}
  \includegraphics[width=\dcollength]{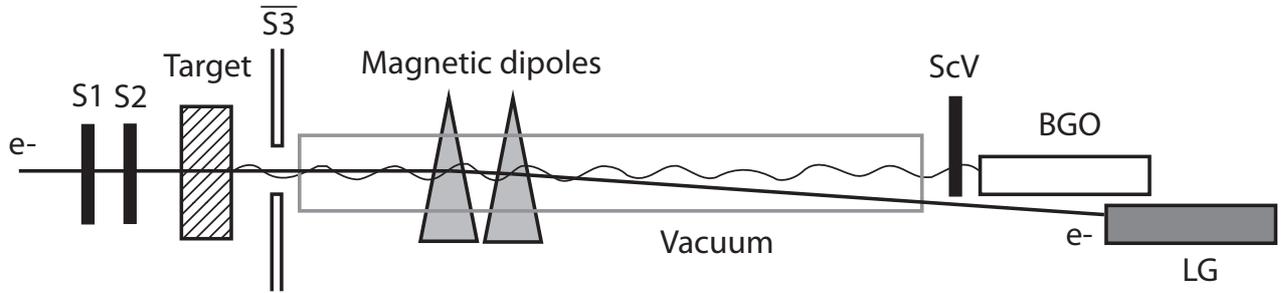}
  \caption{A schematical drawing of the setup used in the experiment. Two magnetic dipoles, each 2 meters along the beam, are used at a low field, $B\simeq0.16$ T, to reduce the influence of synchrotron radiation in the MeV region. This, combined with the necessity of deflecting the electrons outside the BGO calorimeter, into the lead glass calorimeter, forces a long lever arm. Right in front of the BGO, a veto scintillator, \texttt{ScV}, was mounted to reject events where the photon has converted or events where an electron may have interacted with the vacuum chamber, generating a shower. The total length of the setup was about 70 m.}
  \label{fig:setup}
\end{figure*}

A schematic drawing of the setup is shown in \fref{fig:setup}. The electron beam was defined by the scintillator counters
\texttt{S1}, \texttt{S2}, $\overline{\texttt{S3}}$, where the latter is a \diameter{}9 mm hole scintillator (used as a veto). The beam chosen in this way is centered on the target, and, after having traversed this, the electrons are swept away by two dipole magnets, connected in series. This method of separation inevitably introduces Synchrotron radiation (SR), the implication of which will be discussed below. To be able to correct for the background, an empty target run was performed. Vacuum pipes were used when possible, to
reduce background to a minimum in the 70 m long setup. The beam spot
size at the calorimeter position was determined by scanning a \SI{5}{mm} wide scintillator counter in the vertical and hori\-zontal direction in front of the BGO. From these measurements we find a gaussian beam with a standard deviation of $\sigma \approx \SI{12}{mm}$. We can also use this to determine the beam divergence, since the size of the beam is limited by the \diameter \SI{9}{mm} hole scintillator. If it is assumed that the angle made by the direction of the particle and the beam axis is proportional to the distance of the particle from the axis, we calculate a beam divergence of around \SI{0.2}{mrad} for particles just inside the hole scintillator. This is in good agreement with previous measurements with drift chambers \cite{Esbe10,Ande12b}.

The emitted photons are detected by a BGO calorimeter which is described in section \ref{calorimeter}. The electrons that only emit a low-energy photon will be detected by a lead glass (LG) calorimeter which is positioned next to and slightly downstream of the BGO detector. If the electrons have emitted a high-energy photon, the electrons are deflected into the vacuum pipe and not detected. To avoid events where electrons hit the BGO, we have positioned a scintillator (\texttt{ScV}) in front of the BGO. Besides measuring the BGO signal we also measure the height of the signal just before an event. This is motivated in section \ref{calorimeter}.

The data was recorded using an event-based VME system making offline event-selection
possible. The count rate was $\simeq1.6\cdot10^4$ electron triggers
$\mathtt{S1}\cdot\mathtt{S2}\cdot\overline{\mathtt{S3}}$ per burst, with burst duration 9.7 s repeated every 44.4 s, for all targets. 

\subsection{Targets}
All targets consisted of layers of disc-shaped foils with a diameter of
\diameter{}25 mm (except the carbon targets which had a square shape) and different thickness $\delta t$. The number
of layers $N$ in a target was selected such that the total target thickness
$\Delta t = N\cdot\delta t$ would correspond as closely as possible to
$105~\si{\um}$ of tantalum which is $2.56\%~\radl$, where $\radl$ is the radiation length of the target. The choice of total
thickness was a trade-off between
obtaining an acceptable signal-to-background ratio and keeping
multi-photon events at a minimum. The influence of the latter is
already substantial at this thickness. E.g. for \SI{178}{GeV} electrons, the BH yield is reduced by 15\% for a total radiated energy of \SI{1}{GeV}. Keeping
target thicknesses in units of $\radl$ almost constant makes the
contribution from multi-photon events similar in all spectra, as the
correction---essentially a shape-function---should be a polynomial of
the variable $\Delta t/\radl$ \cite{Baie99a}. All targets are listed in \tref{tab:targets}.

The reference target of aluminum was assembled from 80 discs, each of \SI{25}{\um} thickness  and \SI{1}{mm} between the individual foils as described in detail in \cite{Thom10}. Since each
indi\-vi\-du\-al Al foil is significantly thinner than the multiple scattering length of the material, $\lmult=\frac{\alpha\radl}{4\pi}=51.7~\si{\um}$,
this target performs essentially as 'single scatterers' i.e.\ in the BH regime and is a simple, good reference spectrum. The other reference target, tantalum, was chosen since several previous investigations of tantalum are in agreement with theory \cite{Thom09,Thom10}. 

\begin{table}
\setlength{\tabcolsep}{8pt}
	\centering
		\begin{tabular}{r | l l l l l }
		\hline
		\hline
		Material &   $N$   &   $Z$    &$\delta t (\si{\um})$    &   $\radl (\si{cm}) $   &$  \Delta t / \radl$ \\
		\hline
		LDPE    &  13                &    $\sim2$  &     \SI{1}{mm}                        &   50.31                 &         2.58\% \\
		C            &  2                 &     6  &    \SI{2.5}{mm}               &   25.4                   &         1.97\% \\
		Al            &  10              & 13    &          257                          &   8.897               &         2.90\% \\
		Al ref      &  80              & 13    &          25                          &   8.897               &         2.3\% \\
	  Ti           &  9                 & 22    &          106                          &   3.560                 &         2.68\% \\
		Fe         &  6                  & 26    &          80                          &   1.757                 &         2.73\% \\
		Cu     &  5                     & 29      &          74                          &   1.436                 &         2.58\% \\
		Mo    &  5                     &  42   &          49                          &   0.9594                 &         2.55\% \\
		Ta    & 1                     &  73   &         105                          &   0.4094                 &         2.56\% \\
		\hline
		\end{tabular}
		\caption{Specifications of the targets used. From left to right the columns correspond to the material, the number of foils used, the atomic number of the material, the measured thickness of a single foil, the radiation length \cite{PDG12} and the total target thickness in units of radiation length. Notice that $\radl$ for carbon is scaled, since the measured density is significantly different from the tabulated density of PDG \cite{PDG12}.}
		\label{tab:targets}
\end{table}

\subsection{The calorimeters} \label{calorimeter}
The main detector in the setup is the BGO calorimeter which is used to 
measure photons with energies from about \SI{50}{MeV} to \SI{4}{GeV}. In addition, a lead glass calorimeter was used to measure electrons with 
energies above 2 GeV.

\begin{figure}
\centering
     \includegraphics[width=0.7\columnwidth]{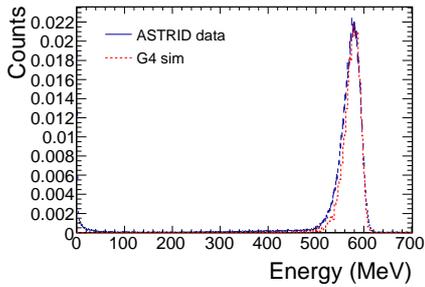}
         \caption{\label{fig:BGO_cal_G4} The BGO spectrum for 580 MeV electrons obtained from ASTRID compared to a Geant4 simulation of the energy deposit in the BGO crystal. }
\end{figure}

\begin{figure}
 \centering
	\includegraphics[width=0.7\columnwidth]{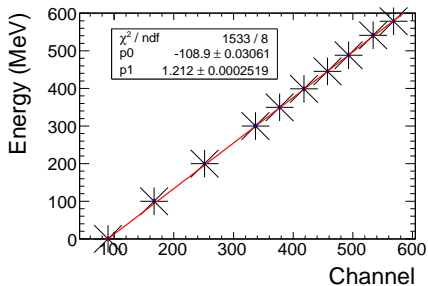}
				\caption{\label{fig:BGO_cal}  BGO calibration made at ASTRID with electrons from 100 MeV to 580 MeV and fitted with a linear function. }
\end{figure}

The BGO detector has a cylindrical shape with \diameter\SI{75}{mm} and 
200 mm length which corresponds to $18X_0$. It is coupled to a Photonics XP3330 
Photomultiplier Tube (PMT) and preequipped with a \SI{12}{V} Scionix preamplifer. 
The detector has been calibrated with an extracted beam from the Aarhus 
STorage RIng Denmark (ASTRID) where electrons 
with energies from 100 MeV to 580 MeV are available. This has been done since it is not possible to get an electron beam with an 
energy below 10 GeV in the North Experimental Area at CERN. It was found that the energy 
deposit in the BGO was in good agreement with Geant4 simulations in this energy regime (see \fref{fig:BGO_cal_G4}) and that the detector response was linear (see \fref{fig:BGO_cal}). The very high $\chi^2$ for the linear fit in \fref{fig:BGO_cal} is mainly caused by a slight non-linearity at energies below 200 MeV and by not including the systematic errors which are difficult to quantify. A high-energy point at \SI{2.6}{GeV} was measured earlier at CERN with tagged electrons \cite{Thom10}. This measurement showed that the calibration at low-energies could safely be extrapolated to multi-GeV energies and combined with the ASTRID data, this shows that the detector response is linear in the energy regime from \SI{200}{MeV} up to \SI{2.6}{GeV}. At energies below 200 MeV the mentioned slight non-linearity shifts the energies by up to 6 MeV. This does affect the conclusions of this experiment which are not energy sensitive on such a small scale. 

				%

Despite the good linearity of the detector, there are several 
difficulties connected to the BGO detector. In the high-energy 
environment at CERN, the detector will occasionally be hit by photons with 
more than 100 GeV. These events are believed to saturate the PMT and 
affect the detector for a substantial amount of time after the event. We have therefore marked events occuring within $50,200$ and \SI{500}{\micro s} after a high-energy event. We have also measured the detector output just before the pulse. This can be used to ensure that the detector has reached its base level, before a new event occurs. 
Early investigations of the detector led to the conclusion that the PMT 
should be run at a very low High Voltage (HV) of 405 V to minimize saturation. This is significantly below 
the recommendations by the PMT supplier, 1140 V, based on calibrations with Cs-137 giving 661 keV photons. Nevertheless, our measurements show that this does not affect the linearity of the detector, which has been tested several 
times. 

We have tested the BGO detector in numerous ways, to find out if it is well-behaved with these non-standard settings. 
The goal is to investigate how a high-energy event affects the detector. We mimic this with a short light pulse from a diode, a laserdiode and an actual pulsed laser. Since the PMT is glued onto the BGO crystal, it is not possible to disassemble the detector and shine light directly onto the photocathode of the PMT. However, it could be accessed from the back by removing an endcap (Photonics XP3330 are no longer available).
%
%
We used a pulsed nanosecond laser to inject a light pulse into the detector. Surprisingly, the response saturated at signal heights well below ones observed at CERN. The cause of this is uncertain. A possible explanation could be that the laser only illuminates part of the photocathode, causing a smaller signal than if the full photocathode is illuminated. We have investigated how low-energy events occuring shortly after a saturation event are affected. This has been done with a second light pulse from a diode which is triggered by the laser. The delay between the laser pulse and the diode was varied from \SI{84}{\micro s} to \SI{400}{\micro s} and only a small effect was observed, with a reduction in the signal height of about $\sim10\%$ when the delay was reduced from \SI{260}{\micro s} to \SI{84}{\micro s}. No effect was seen when the delay was reduced from \SI{400}{\micro s} to \SI{260}{\micro s}. Since the affected data are flaged by the DAQ system, this effect can be removed from the data in the analysis.

\begin{figure}
\begin{minipage}[b]{0.48\linewidth}
\centering
     \includegraphics[width=\textwidth]{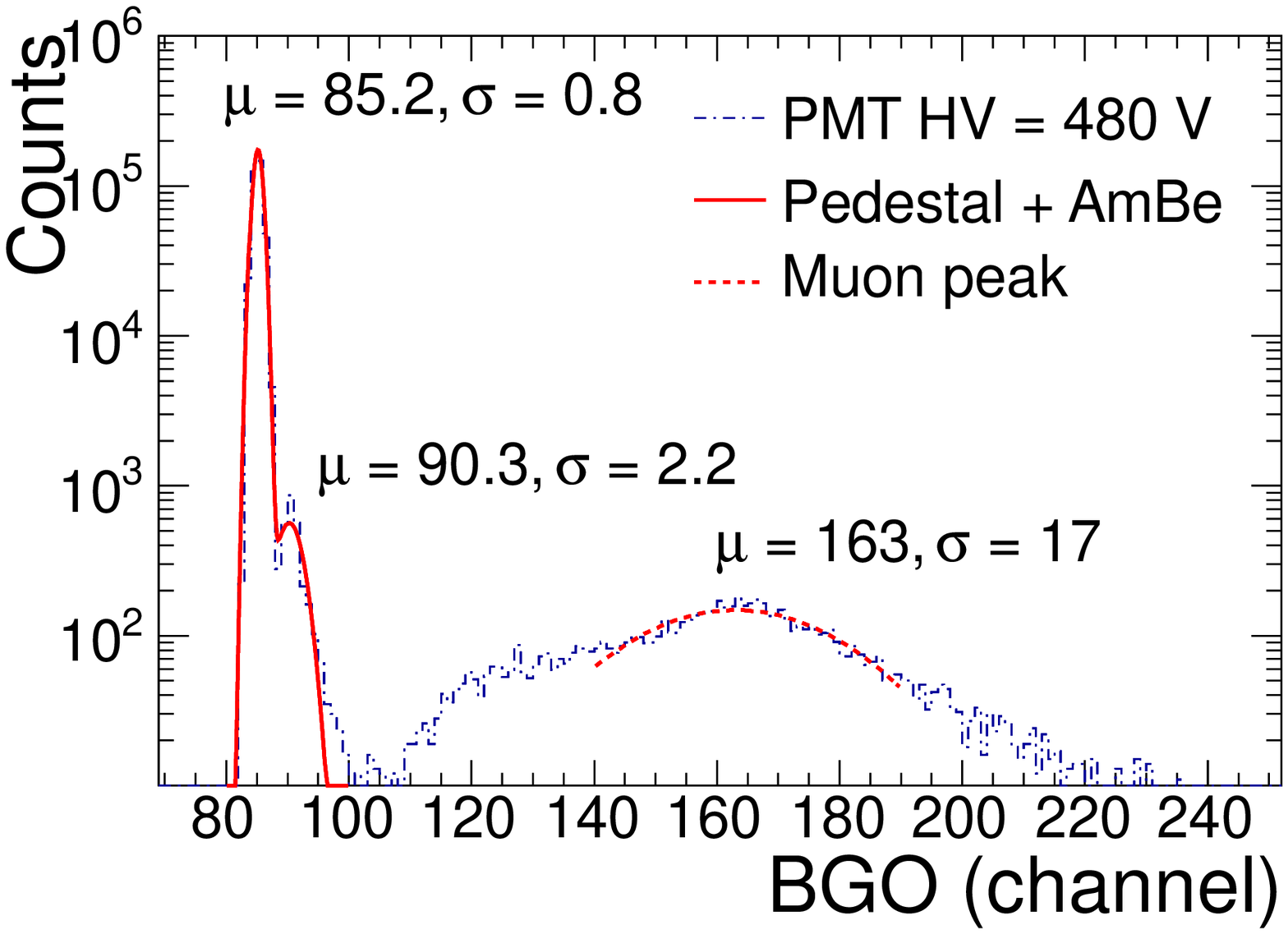}
     \end{minipage}
\hfill
     \begin{minipage}[b]{0.48\linewidth}
\includegraphics[width=\textwidth]{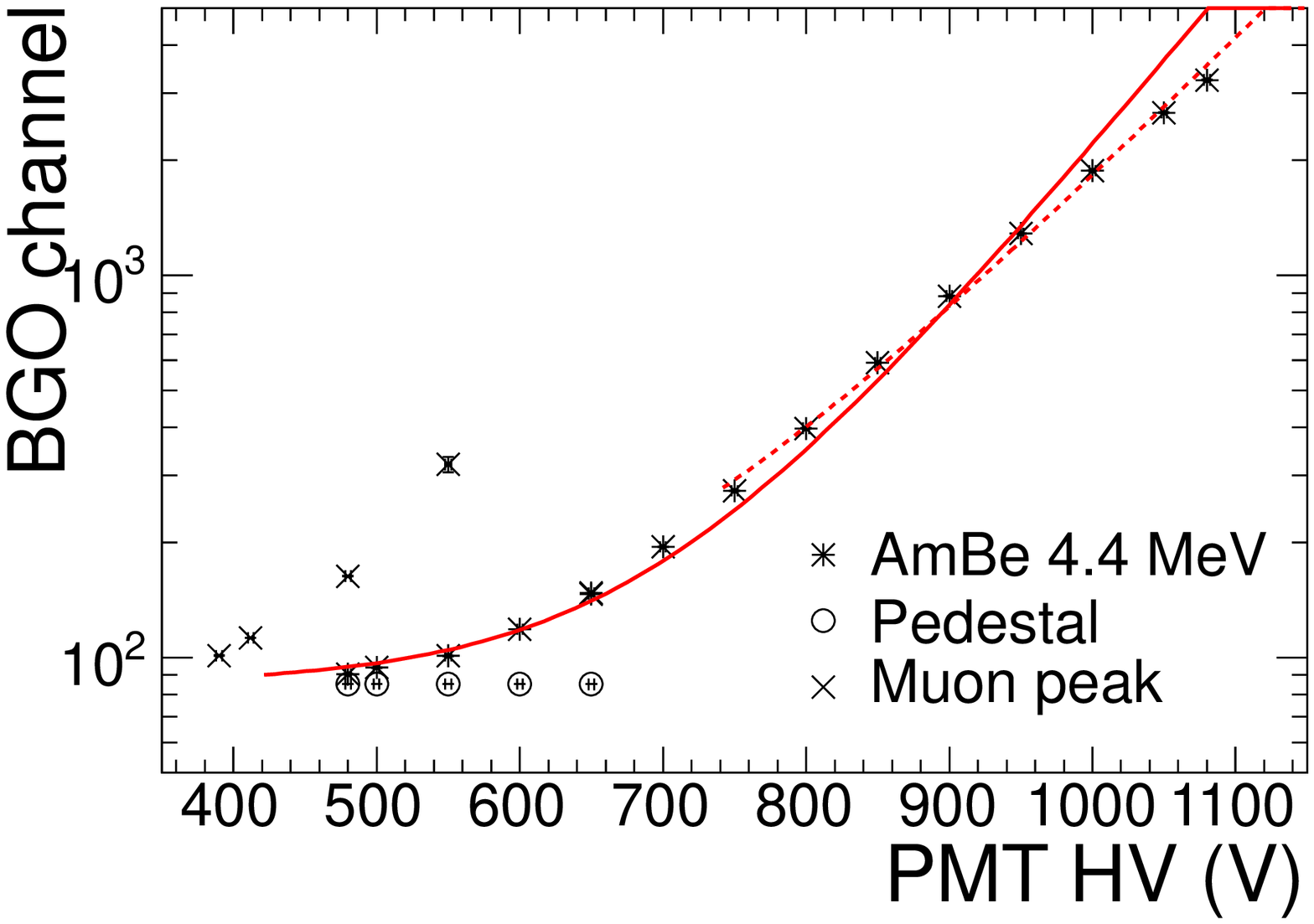}
         \end{minipage}
         \begin{minipage}[t]{0.48\linewidth}
         \centering
         \caption{\label{fig:AmBe} The BGO spectrum measured at a PMT HV 
of 480 V. One can observe the pedestal at channel 85, a 4.4 MeV peak 
from the AmBe source, and a broad muon peak from cosmics.}
          \end{minipage}
         \hfill
         \begin{minipage}[t]{0.48\linewidth}
             \caption{\label{fig:AmBepeak} The peak channel as a 
function of PMT HV for the 4.4 MeV peak, the muon peak and the pedestal. 
The AmBe data have been fitted to a function describing the PMT 
amplification in two intervals. }
     \end{minipage}
     \end{figure}

We have investigated the signal height as a function of the HV with a AmBe source giving 4.439 MeV photons. The spectrum measured at a HV of 480 V is plotted in \fref{fig:AmBe}. One can see the pedestal at channel 85, the 4.4 MeV 
peak at channel 90, and a broad muon peak around channel 163. The muon peak 
is caused by cosmic muons that deposit up to 100 MeV in the crystal. 
This peak is broad since the detector is cylindrical and the particles 
therefore traverse different thicknesses of material. The positions of the 4.4 MeV peak, the muon peak, and the pedestal are shown as a function of the HV in \fref{fig:AmBepeak}. As a simple model of the BGO signal as a function of the PMT HV we use
\begin{equation}
f_{PMT}(\text{HV}) = a_{ped}  + c\cdot b^{\text{HV}}.
\label{eq:fPMT}
\end{equation}
$a_{ped} = 85$ is the pedestal level and $b$ and $c$ are fit parameters. 
For $\text{HV}>\SI{700}{V}$ there is a very good agreement with a
signal doubling every \SI{81}{V}. If the lower voltages are included in the fit,
the agreement is not as good. Nevertheless, the observation of the AmBe signal and the muon peak at low HV
indicates that the detector is well-behaving, even when operating at a HV significantly below the recommended value.

In conclusion we have tested the BGO detector in various different ways 
and we did not find any critical effects even though the HV setting for the 
PMT is substantially lower than the recommendation by the supplier. 


\section{Simulations}\label{simulations}
\begin{figure}
\centering
\includegraphics[width=\columnwidth]{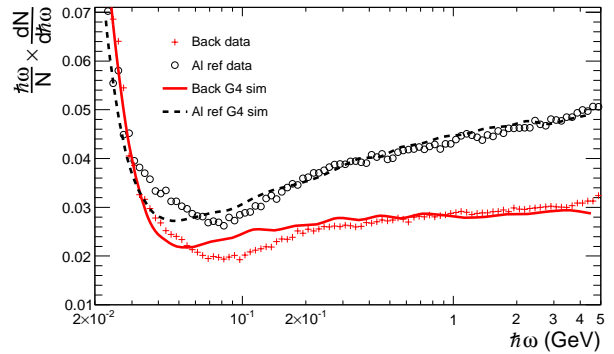}
\caption{\label{fig:G4sim} Geant4 simulations and measurements of the background radiation and the aluminium reference target. See the text for more details.  }
\end{figure}

To compare our measurements to theories we have to make Monte Carlo simulations, since our spectra are affected by multi-photon events, where more photons are emitted by the same electron and detected as one. Such events are easily included in Monte Carlo simulations. Full Geant4 simulations of the background and of the background plus the reference target are shown with the corresponding data in \fref{fig:G4sim}. In this simulation we include the full geometry of the setup, use a gaussian beam divergence with $\sigma= \SI{0.225}{mrad}$, a gaussian beam distribution with $\sigma = \SI{10}{mm}$ and include a background according to our measurements. To find a good agreement between the simulations and the data we have added an extra background of $1.7\% \radl$ which is probably caused by two beam control scintillators positioned $\SI{14}{m}$ upstream the targets (not shown in \fref{fig:setup}). In the simulation we store the energy deposit in the BGO, in the lead glass and in the veto scintillator placed in front of the BGO. We remove all events where more than \SI{0.5}{MeV} is deposited in the veto scintillator and take into account that only 92\% of the particle energy is deposited in the BGO. This fraction has been determined from simulations with mono-energetic beams. 

The simulation results for the background and the aluminium reference target are shown in \fref{fig:G4sim}. For the data we have applied a cut removing all events with a hit in the veto scintillator and all events within \SI{500}{\micro s} from a high-energy event. The simulations and the data agree very well above \SI{200}{MeV} and for the low-energy rise below \SI{30}{MeV}. The magnetic field of the two dipoles is \SI{0.17}{T} which introduces SR with a critical energy of \SI{3.6}{MeV} which is well below the low-energy rise. From simulations with and without the LG calorimeter it is clearly seen that the main cause of the low-energy rise is backsplash from the LG. This arises since events where only low-energy photons are emitted, are accompanied by a high-energy electron which is only deflected $\sim\SI{7}{cm}$ by the two dipoles at the position of the detectors. Unfortunately, the amount of backsplash was not discovered during the experiment. 
In the intermediate energy region from \SI{30}{MeV} to \SI{200}{MeV} there is a dip in the radiation spectrum which is not well understood. Several explanations for this dip have been tested, but all unsuccessful. 

\begin{figure}
\centering
\includegraphics[width=\columnwidth]{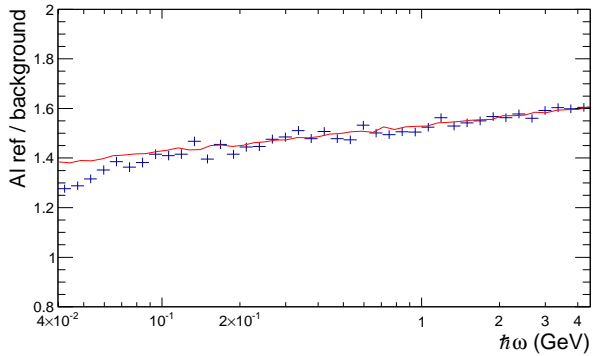}
\caption{\label{fig:BackRefRat} The measured (markers) and simulated (line)ratio between the aluminium reference and the background.}
\end{figure}

\begin{figure}
\centering
\includegraphics[width=\columnwidth]{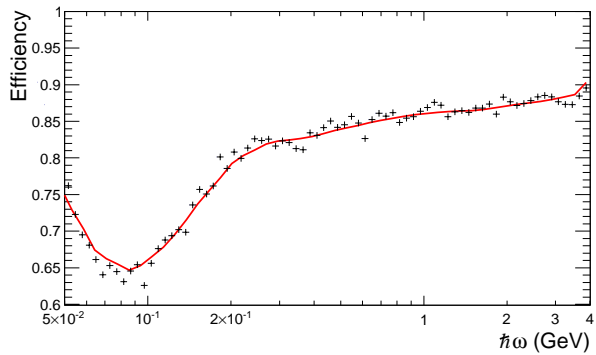}
\caption{\label{fig:eff} Detection efficiency found by dividing the measured radiation background to the simulated as described in the text. The markers are the actual ratio and the line is the smoothed efficiency used in the analysis.}
\end{figure}

Since it has not been possible to accurately reproduce the measured spectrum in the full energy range with the Geant4 simulations (although it is rather close), we have chosen another approach to simulate the radiation spectra, which is significantly simpler and more physical transparent.  This method has been used in a previous experiment \cite{Ande12a} and only calculates the radiation emitted by the electron and does not track the particle through the setup. It includes a BH radiation background, SR from the magnets and radiation from the target. We ignore the energy loss of the electron, which can safely be done since we are only interested in events with a low energy loss. However, since we do not track the particle through the setup and therefore do not include the veto scintillator, we have to determine a detection efficiency. This was not neccessary for the Geant4 simulations above. We assume that the efficiency can simply be modelled as a multiplicative factor which varies slightly with energy. We use the ratio of the radiation spectrum from the aluminium reference to the background to determine the amount of material in the background (see \fref{fig:BackRefRat}). We use these two, since both only include SR and BH radiation, which are well-known. By ta\-king the ratio between the two measurements, we remove the detection efficiency, and we can directly compare our data to the simulations. We find that the amount of background material is $3\%\radl$, which corresponds to the $1.7\%\radl$ mentioned earlier plus scintillators, mylar foil, air, etc. in the target area. With this background we find the efficiency by dividing the background measurements with the simulations. The result is shown in \fref{fig:eff}. We observe a significant efficiency drop from \SI{50}{MeV} to \SI{200}{MeV} as was also expected from the Geant4 simulations, but from the ratio (see \fref{fig:BackRefRat}) it is reasonable to assume that the efficiency corrected simulations are valid down to $\sim\SI{100}{MeV}$, where also the simulations begin to deviate from our measurements.  For the \SI{20}{GeV} data we have used the same method, except that SR is neglected since the critical energy is \SI{0.1}{MeV} and therefore way below our detection threshold.

\subsection{Target Radiation}
For the radiation from the target we use the Migdal \cite{Migd56} cross section calculated with the approximations of Stanev et al. \cite{Stan82}. This is considered a standard within the field. However, after the SLAC measurements several other theoretical models were developed. For \SI{178}{GeV} electrons in carbon and \SI{20}{GeV} in copper, we also calculate the radiation cross section according to the formalism of Blankenbecler \& Drell (BD) \cite{Blan96}. Blankenbecler later refined their treatment of the problem and included an extra correlation term \cite{Blan97b}. We have calculated both expressions for the two cases. Alternative descriptions of the LPM effect have also been developed by Baier \& Katkov (BK) \cite{Baie98b} and Zakharov \cite{Zakh96b}. The BK description has been calculated for 178 GeV electrons in carbon.

\begin{figure}
\centering
\includegraphics[width=\columnwidth]{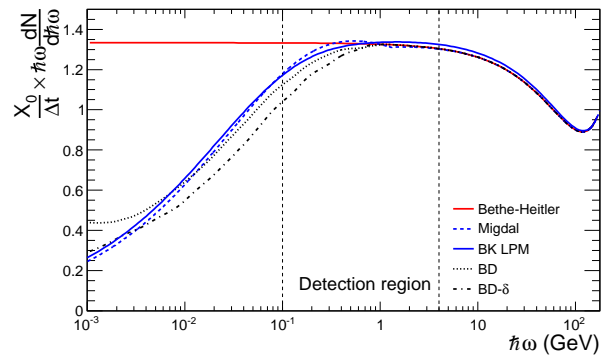}
\caption{\label{fig:crosssections} Calculated radiation power spectra for 178 GeV electrons  traversing a $1.97\%\radl$ carbon target. The radiation spectra are normalized to the target thickness in units of $\radl$. The Migdal line is valid for a semi-infinite target and calculated with the approximations of Stanev et al. BK refers to the theory of Baier and Katkov. The Blankenbecler and Drell curves (BD and BD$-\delta$) are calculated for a finite size target. The $\delta$ refers to formulas in \cite{Blan97b} including an extra correlation term.  }
\end{figure}

\begin{figure}
\centering
\includegraphics[width=\columnwidth]{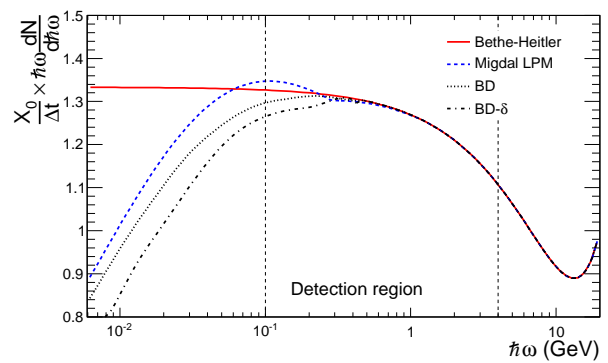}
\caption{\label{fig:crosssections20} Calculated radiation power spectra for 20 GeV electrons  traversing a $2.58\%\radl$ copper target. See caption of \fref{fig:crosssections} for more info. }
\end{figure}

The spectra of the radiated power by \SI{178}{GeV} electrons traversing a carbon target with a thickness corresponding to $1.97\%\radl$ is shown in \fref{fig:crosssections}. The differences between the curves are small but worth mentioning. For the BD curve we observe a flattening around 5 MeV. This is related to the finite thickness of the target. If the target thickness is equated to the formation length given by $l_f = 2\gamma^2c/\omega$ one finds $\hbar\omega = \SI{10}{MeV}$. This means that for photons with an energy below \SI{10}{MeV}, one cannot consider the target as semi-infinite and one has to take the size into account. This is naturally incorporated into the BD formalism, but not included in Migdal's formulas. However, this is below our detection limit and does not affect our measurements. We can also ignore transition radiation and dielectric suppression for the same reason. For an experimental investigation of the effects related to the finite target size in the context of bremsstrahlung, the reader is directed to \cite{Thom10}. It is also interesting to notice that it is only the Migdal formulation which shows a small bend around \SI{1200}{MeV}. The BK curve does not show this but is smooth and otherwise very close to the Migdal curve. The BD curve is also very similar to Migdal curve above \SI{10}{MeV}, but the BD$-\delta$ curve is consistently slightly below except for the lowest energies, where the size of the target plays a role and one cannot directly compare them. 
The 20 GeV spectra in \fref{fig:crosssections20} generally show the same tendencies and the same tendencies are seen for other target materials.

\begin{figure*}[ht]
\centering
\includegraphics[width=0.33\textwidth]{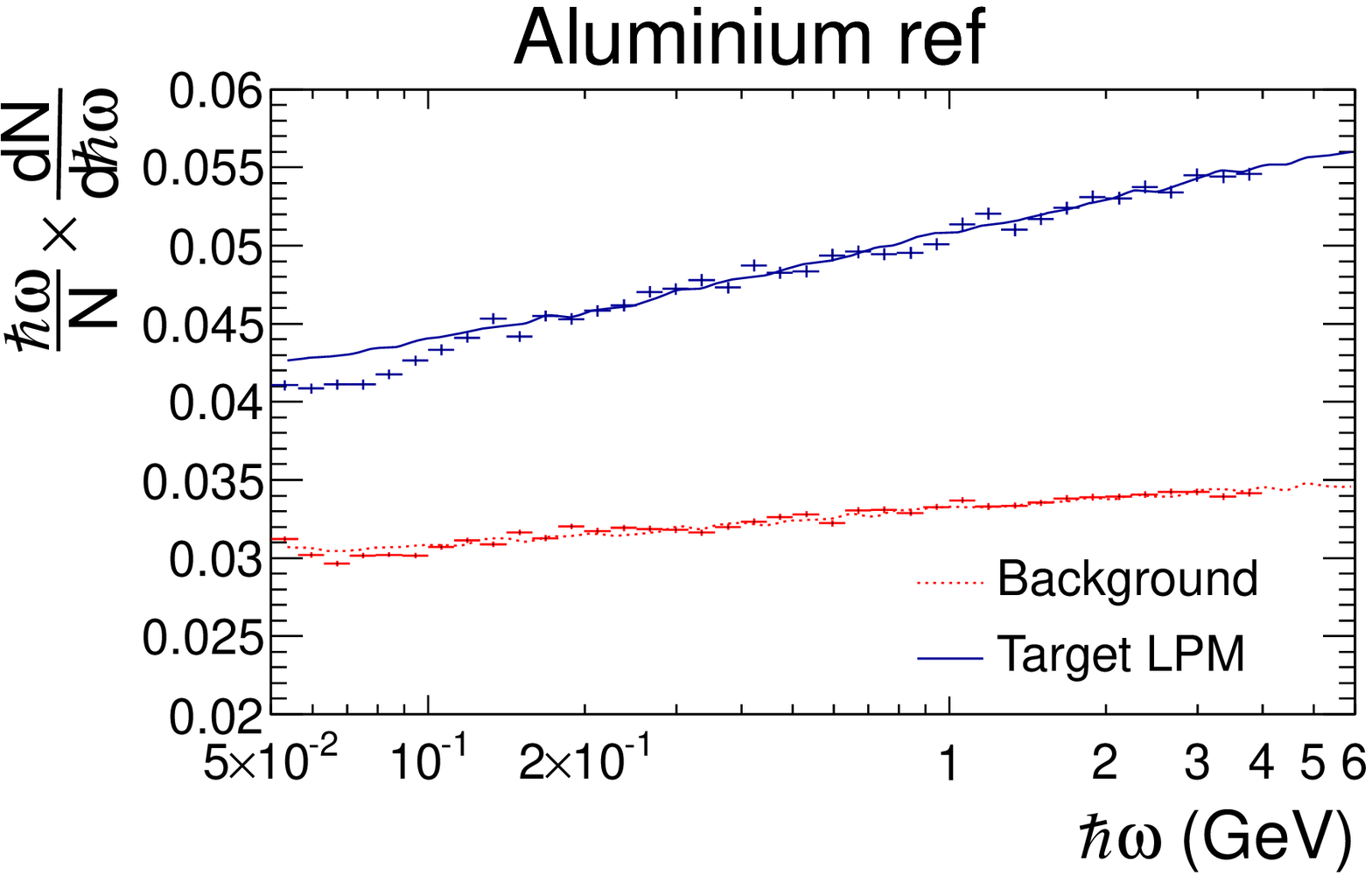}
\includegraphics[width=0.33\textwidth]{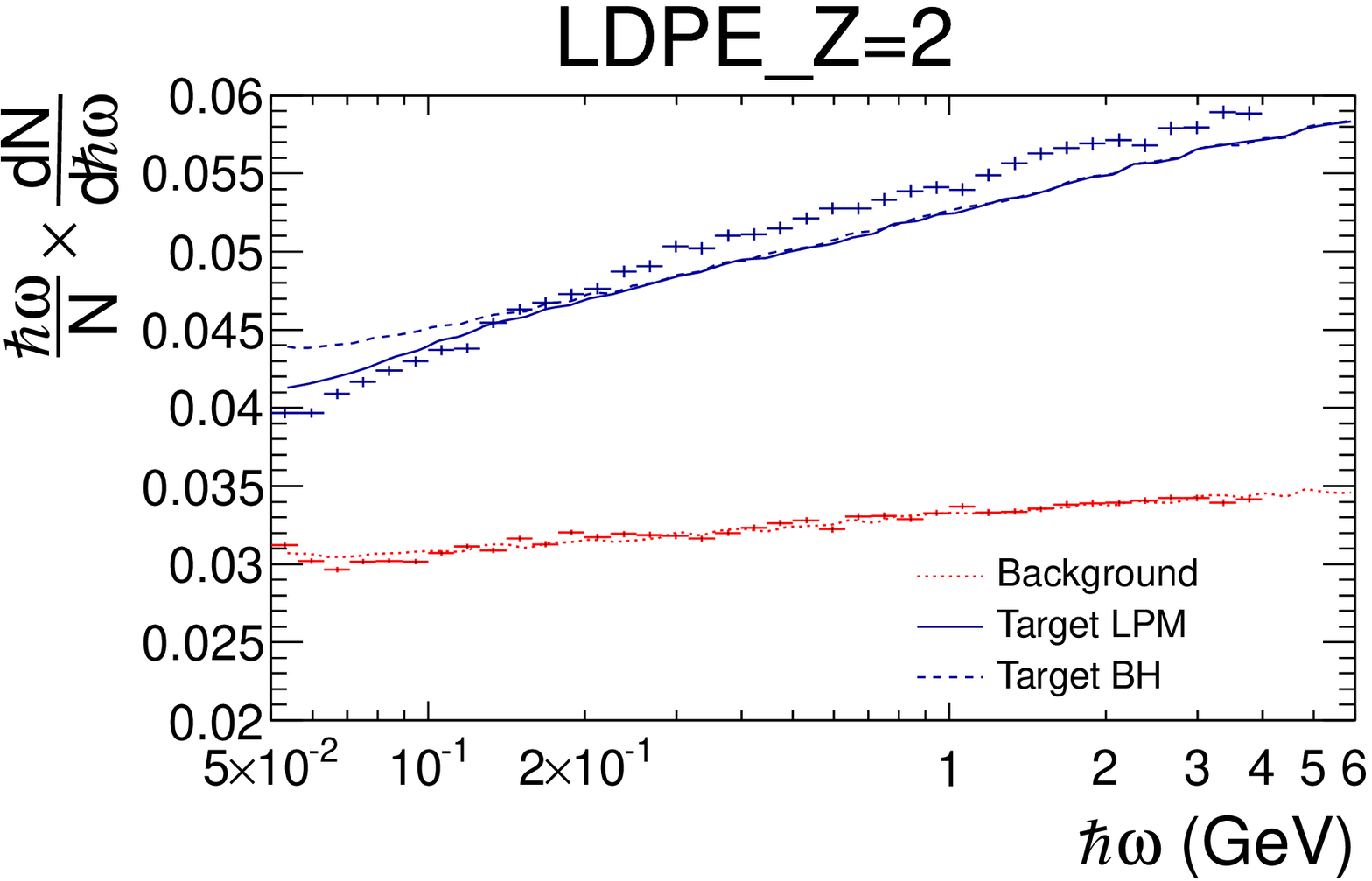}
\includegraphics[width=0.33\textwidth]{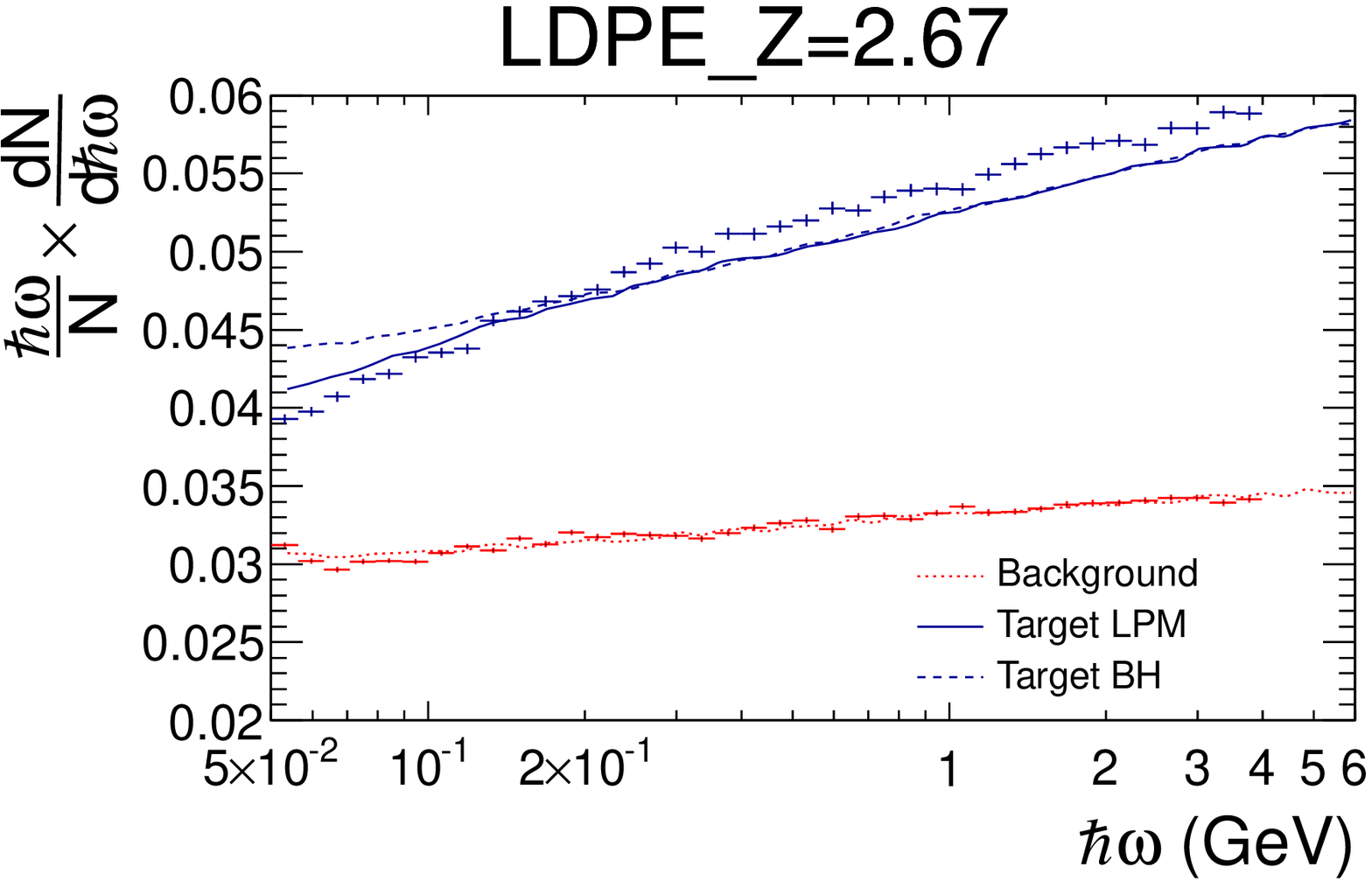}
\includegraphics[width=0.33\textwidth]{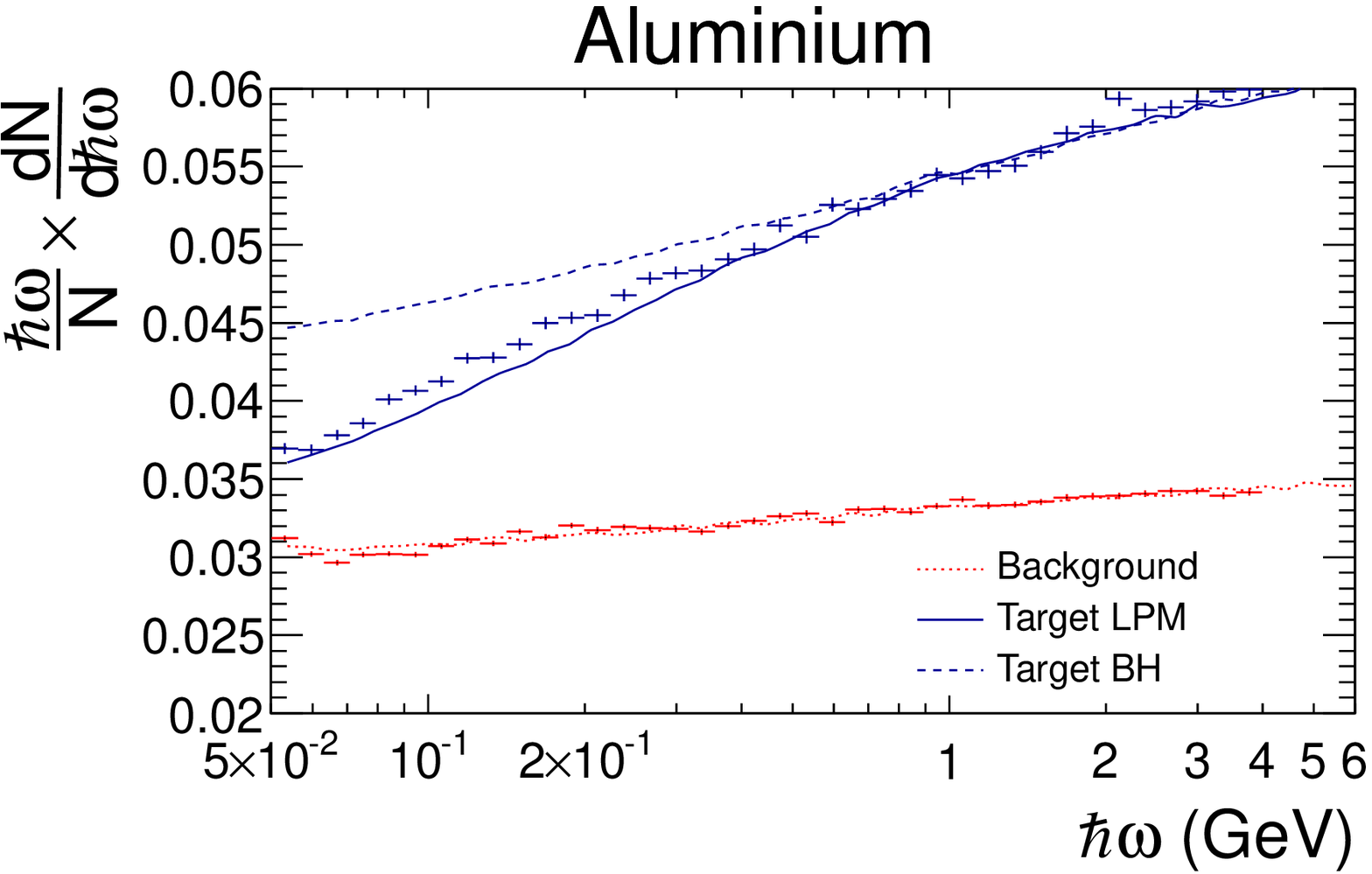}
\includegraphics[width=0.33\textwidth]{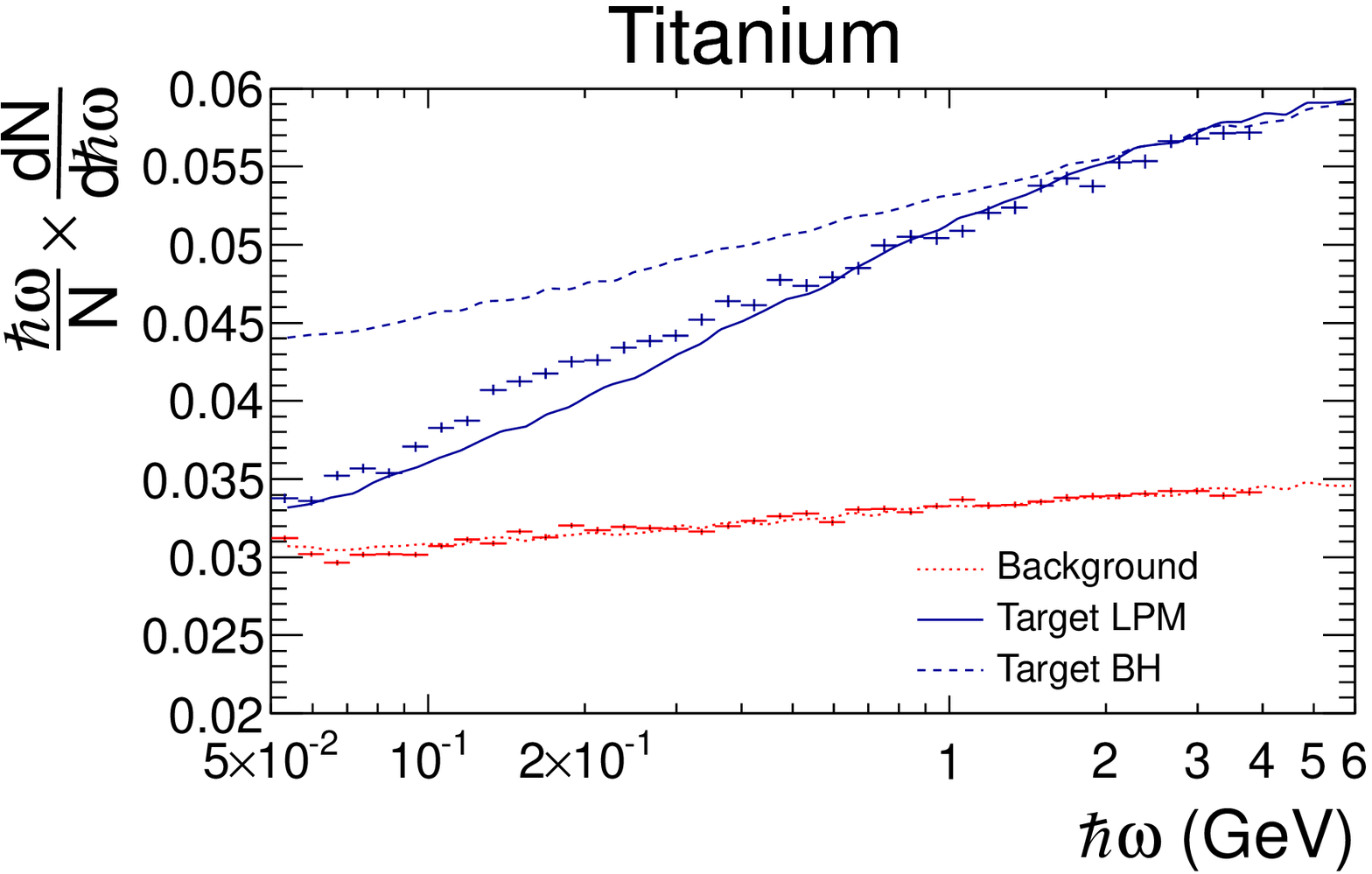}
\includegraphics[width=0.33\textwidth]{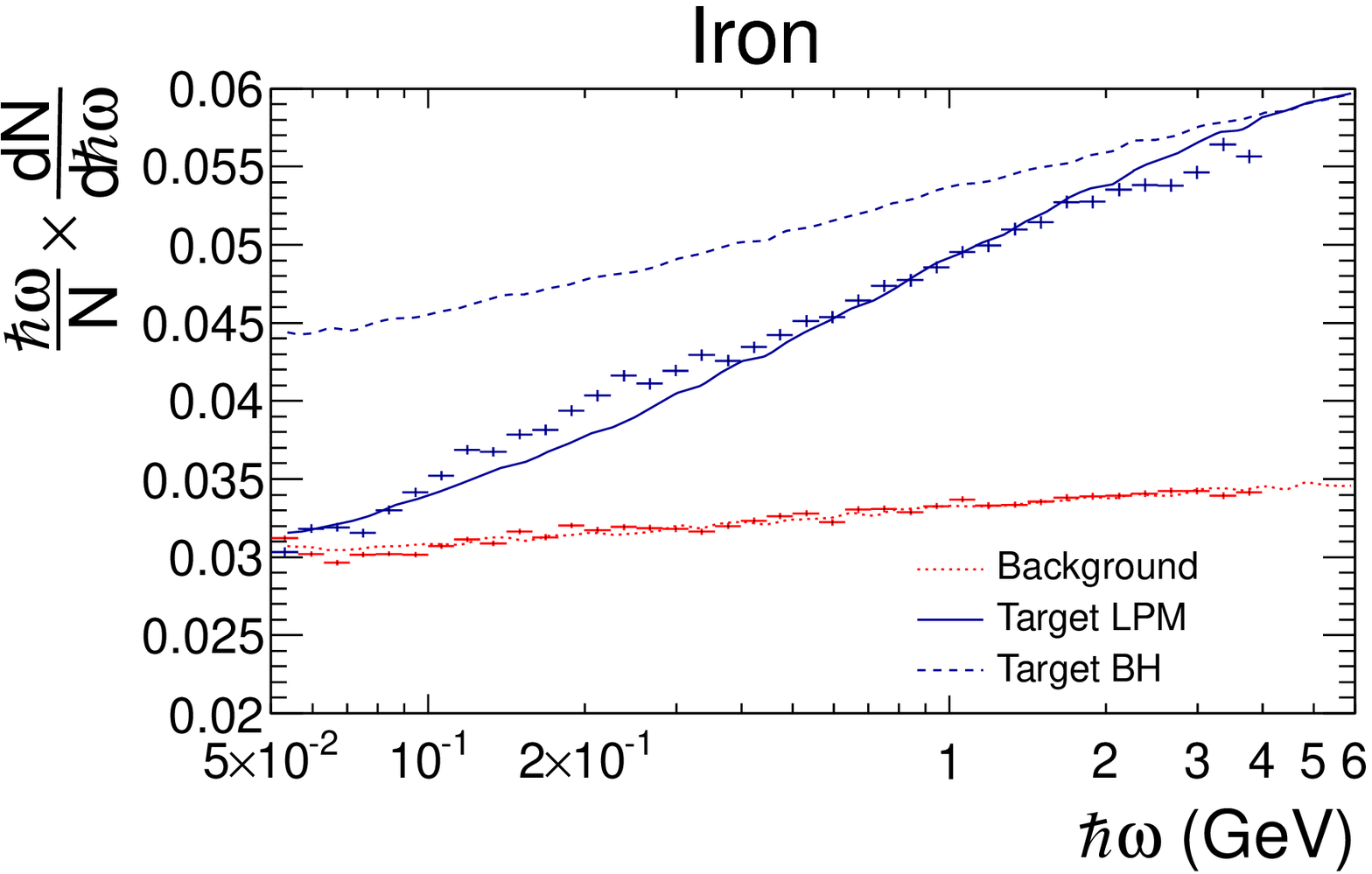}
\includegraphics[width=0.33\textwidth]{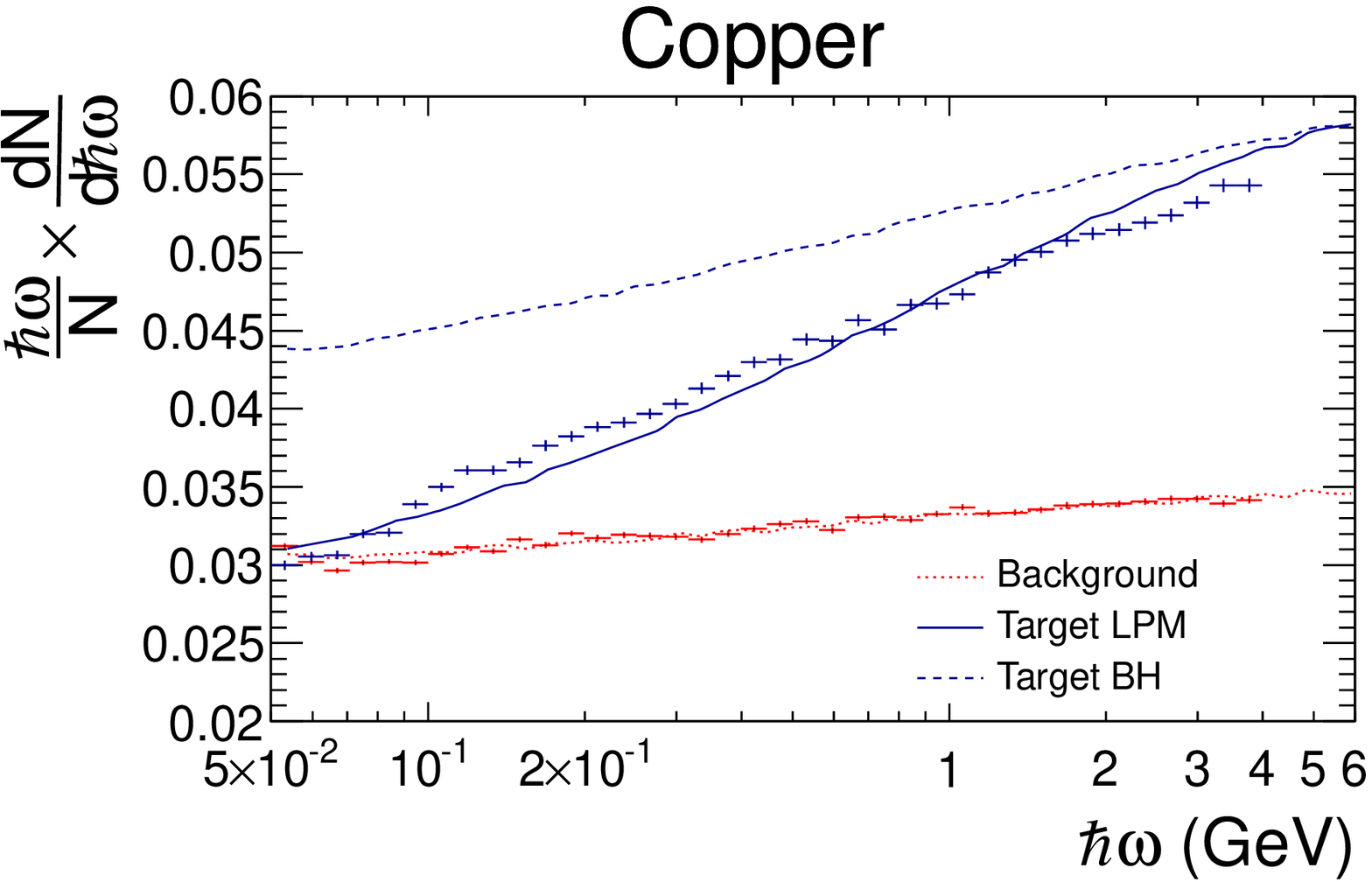}
\includegraphics[width=0.33\textwidth]{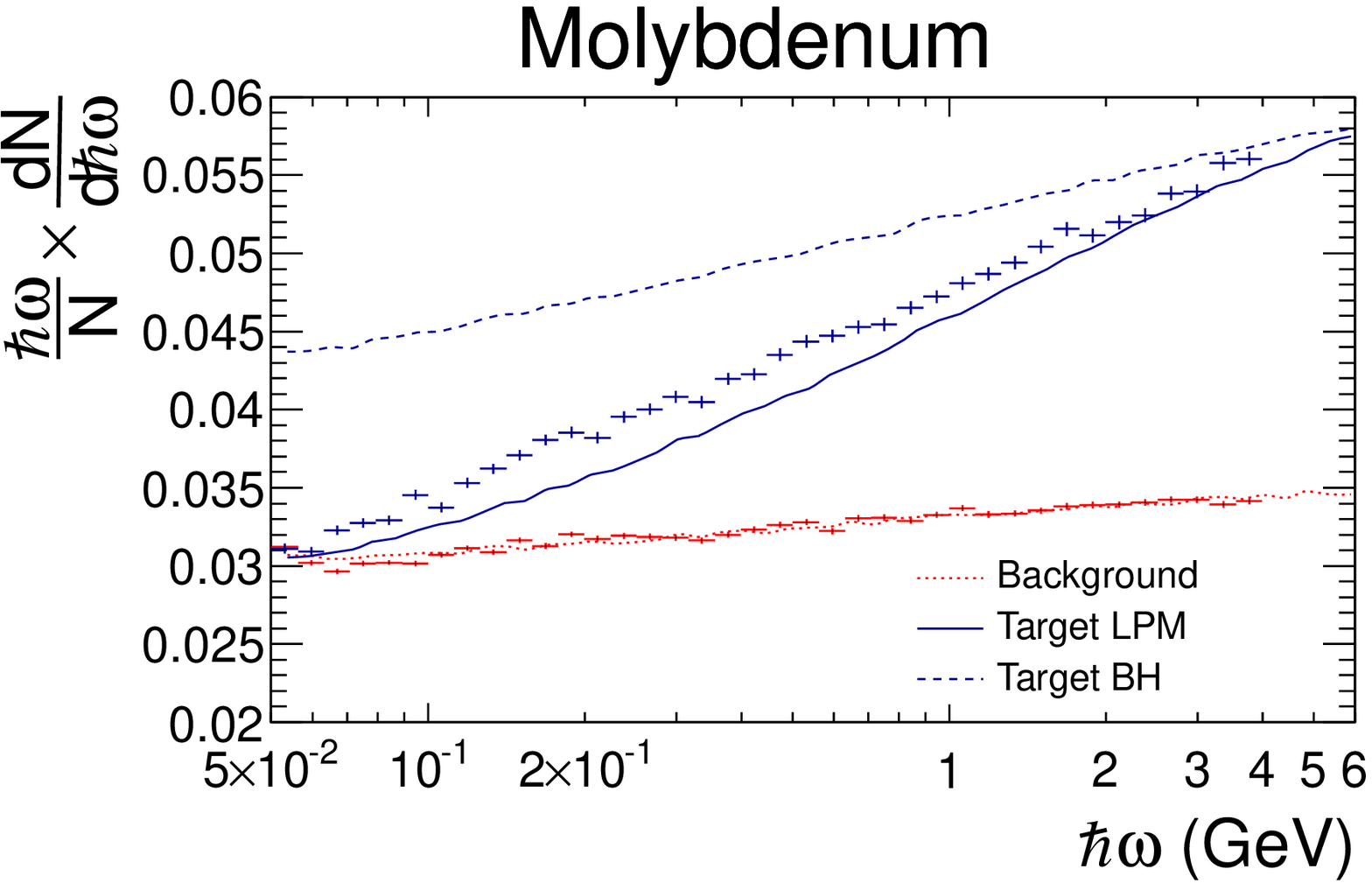}
\includegraphics[width=0.33\textwidth]{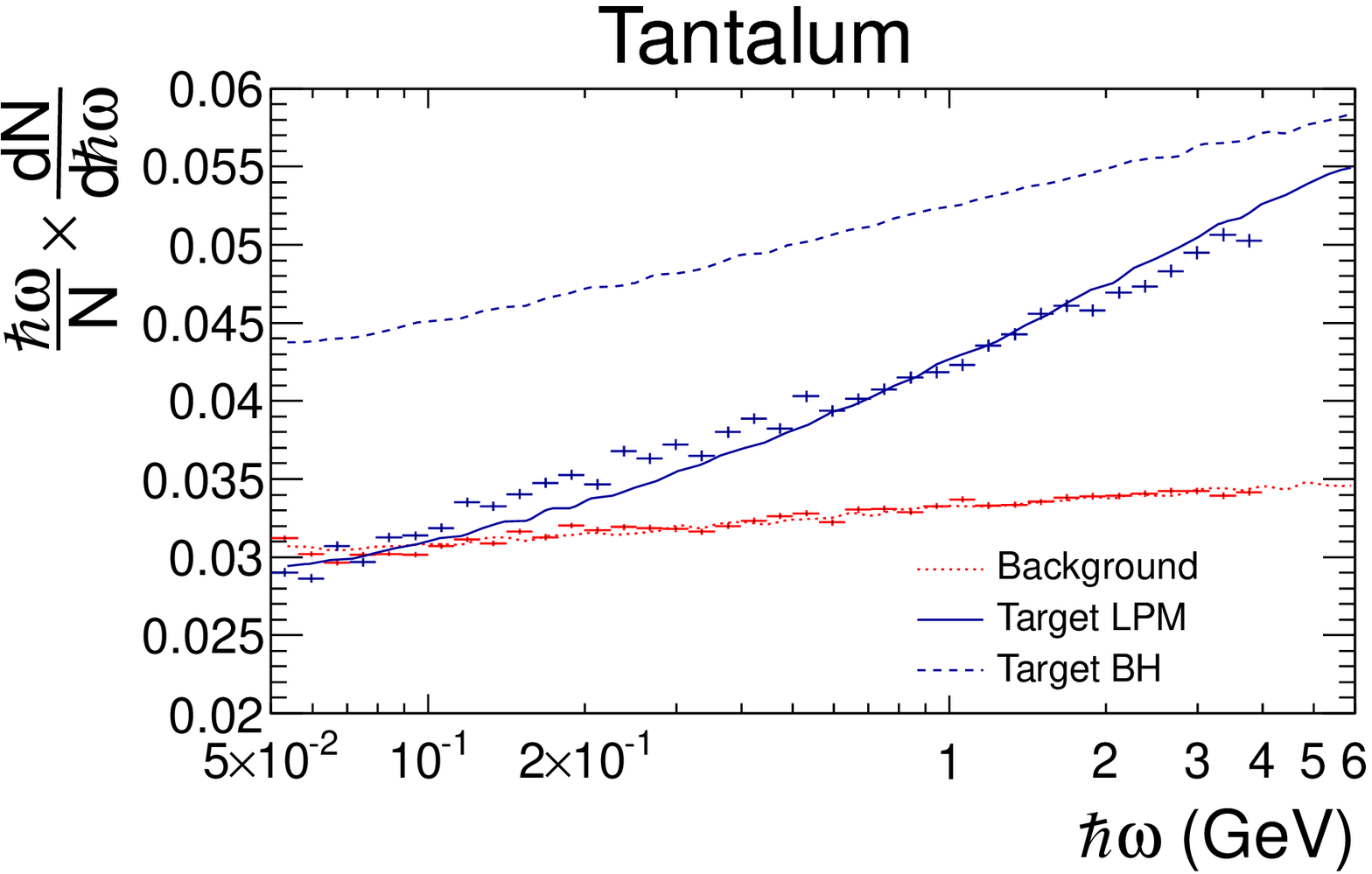}
\caption{\label{fig:results1}Power spectra of radiation emission from 178 GeV electrons penetrating amorphous targets. The spectra are normalized to the number of incoming electrons. The lower, red dotted line shows the simulated contribution from the background, with data points representing the values obtained in the experiment. The middle, blue line shows the simulated contribution from the targets, including the LPM effect, and with data points representing the values obtained in the experiment. The upper, black dashed line shows the simulated contribution from the targets, excluding the LPM effect. Only statistical uncertainties are shown.  }
\end{figure*}

\section{Results and discussion}

Our results obtained with 178 GeV electrons are shown in \fref{fig:results1}. For the reference targets of aluminum and tantalum there is generally a good agreement between data and simulated values (for Ta only when including the LPM effect, as expected), except perhaps for the lowest (intermediate) photon energies 50-120 (120-600) MeV for the Al (Ta) target where a $\simeq5\%$ discrepancy is seen. The discrepancy below 100 MeV may be attributed to influence from the synchrotron radiation and/or backsplash in the BGO as mentioned in section \ref{simulations}.
For the LDPE target we assume the composition is CH$_2$ and have calculated curves based on the average $Z$ value of 8/3 and the average RMS $Z$ value of $\sqrt{6^2 + 2\cdot 1^2}/3 \simeq 2$. The latter corresponds to a $Z^2$-scaling of the main radiation mechanism.

These are hardly distinguishable. Above \SI{200}{MeV} our data are consistently above the simulations which suggests that the target is thicker or of a different composition than assumed. There is also a slight tendency for the data to be a steeper function of photon energy than the simulated values. For aluminum and titanium there is again good agreement with si\-mu\-lations including the LPM effect, except for a small systematical shift of experimental values below about 400 MeV. For iron and copper targets - with very similar atomic numbers - the spectra are close to being identical with an indication of a change of slope at the $\simeq5\%$ level. Finally, for the medium $Z=42$ molybdenum, the data points are consistently 5-10\% higher than the simulated values including the LPM effect, indicating a systematic error.

\begin{figure}
\centering
\includegraphics[width=\columnwidth]{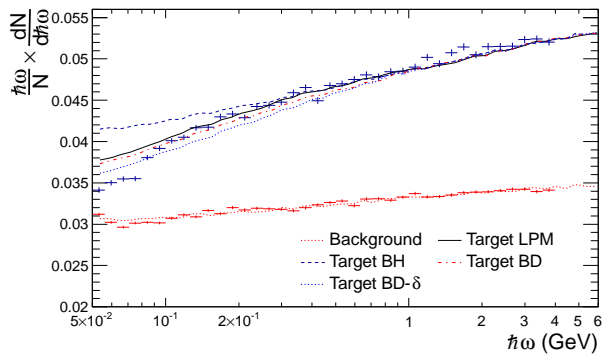}
\caption{\label{fig:resC} 178 GeV electrons in carbon target. See caption for \fref{fig:results1}. Furthermore is included simulations based on the BD and BD$-\delta$ theory. }
\end{figure}

For carbon (see \fref{fig:resC}) we have also calculated the spectrum for the BD and BD$-\delta$ theory. In the energy interval from 100 MeV to 1 GeV where the theories differ we have calculated the $\chi^2$ value. The number of degrees of freedom is 20 and $\chi^2_\text{Migdal} = 28,\ \chi^2_\text{BD} = 69$ and $\chi^2_{\text{BD}-\delta} = 209$. In other words the data have a preference for the Migdal formula. There is a slight disagreement with the BD curve and the BD$-\delta$ curve is consistently below our data. 

\begin{figure}
\centering
\includegraphics[width=\columnwidth]{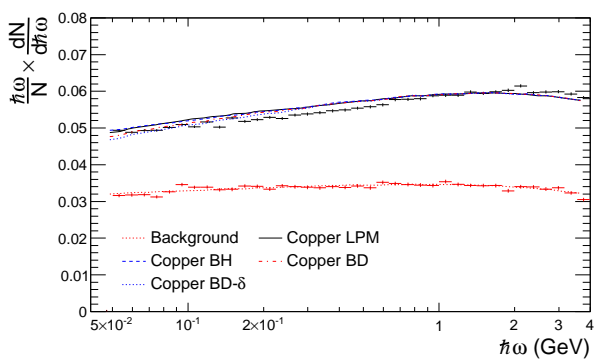}
\caption{\label{fig:res20}  20 GeV electrons in copper target. See caption for \fref{fig:results1}. Furthermore is included simulations based on the BD and BD$-\delta$ theory. }
\end{figure}

Our results obtained with 20 GeV electrons in copper are shown in \fref{fig:res20}. The appearance of a 'kink' in the 'raw' spectrum (see \fref{fig:crosssections20}), disappears when taking emission of multi-photons and background into account. Thus, an investigation of the potential limitations of the Migdal model in this regime requires very thin targets and a very low background. Our experimental values are generally of the same shape as the simulated spectra, but about 4\% lower in magnitude. The reason for this discrepancy is not known, but could be related to contamination of the 20 GeV beam with particles heavier than electrons. Previous measurements have shown that a contamination by non-radiating particles at the 10-20\% level for low energies ($\lesssim\SI{25}{GeV}$, where synchrotron radiation in the main bends becomes insignificant) can be present if the beam is not carefully tuned. The data cannot be said to have a pre\-fe\-rence for any of the theoretical models calculated, but are in agreement with all.

\section{Conclusion}
We have investigated the LPM effect for 20 and 178 GeV electrons penetrating media of primarily low atomic number. On the scale of about 5\% there is no experimentally based reason to suspect that the widely applied model of Migdal should be inadequate, in spite of the statistical nature of the screening function applied and the differing LPM contribution from electrons. Thus, for instance simulations of extended air showers based on the Migdal formulation should be reliable - for this mechanism and given that the present test can be extrapolated to higher energies - to an accuracy of better than 5\%. The carbon data have a preference for the Migdal formula, and the BD$-\delta$ curve is consistently slightly below our data. This is in disagreement with previous results on structured targets \cite{Ande12a}, which had a slight preference for the BD$-\delta$ theory. Oppositely, it is in agreement with measurements on thin targets, where the BD$-\delta$ theory was also disfavoured \cite{Thom10b}. 



%

\end{document}